\documentstyle[prl,aps,preprint]{revtex}
\begin{document}
\draft
\tighten

\title{On the origin of the quantum-critical transition in the bilayer
Heisenberg model.}
\author{C. N. A. van Duin and J. Zaanen}
\address{Institute Lorentz for Theoretical Physics, Leiden University\\
P.O.B. 9506, 2300 RA Leiden, The Netherlands}
\date{\today ; E-mail:cvduin@lorentz.leidenuniv.nl; jan@lorentz.leidenuniv.nl}
\maketitle

\begin{abstract}
The bilayer Heisenberg antiferromagnet is known to exhibit a quantum-critical
transition at a particular value of the inter-layer coupling. 
Using a new type of coherent state, appropriate to the special 
order parameter structure of the bilayer, we map the problem onto the
quantum non-linear sigma model. It is found that the bare coupling constant 
diverges at the classical transition of Chubukov and Morr, so that in any finite
dimension the actual transition occurs inside the ordered phase of the
classical theory.
\end{abstract}
\pacs{75.10.-b, 75.10.Jm, 64.60.-i, 75.45.+j}

\narrowtext
The study of non-classical collective quantum states of matter is a
central theme of modern condensed matter physics. Despite the successes in 1+1 
dimensions, it has proven difficult to address these matters in higher 
dimensions. Either the minus-sign problem intervenes (as in, e.g., the t-J model and frustrated spin models), or the tendency towards classical order is too 
strong (e.g., unfrustrated spin models). The class of bilayer Heisenberg models 
is special in this regard\cite{bigen,Hida1}. 
It is sign-free, and convincing numerical evidence exists showing that its 
long-wavelength behavior is governed by the $O(3)$ quantum non-linear sigma 
model (QNLS) with tunable bare coupling constant $u$
 \cite{QMC}. The relationship between the microscopic model and its long 
wavelength behavior is non-trivial. Chubukov and Morr (CM) made the key 
observation that, in order to construct the classical limit, the severe local 
(interplanar) fluctuations have to be integrated out first \cite{CM}. In the 
resulting singlet-triplet representation, a phase transition between a N\'eel
state and an incompressible state is found already at the classical level. 
CM conjectured that this 
transition corresponds to the quantum critical transition found in numerical 
studies. Here it is shown that this is not correct. Because of the special
structure of the order parameter, the standard $su(2)$ generalized spin 
coherent state does not suffice for the construction of the path integral.
We introduce a novel type of coherent state which allows us to straightforwardly
recover the QNLS describing 
the long wavelength behavior. We find that the bare coupling constant of the 
field theory {\em diverges at the classical transition} of CM. The quantum phase transition therefore occurs well before the classical transition can occur, and 
the latter is therefore in any finite dimension an artefact. 

It is convenient to consider the ``bilayer'' model in arbitrary dimensions,
with an added  magnetic field ($\vec{B}$),
\begin{eqnarray}
{\cal H} & = & J_1 \sum_{<ij>}(\vec{s}_{i1}\cdot\vec{s}_{j1} +\vec{s}_{i2}\cdot\vec{s}_{j2})+ \nonumber \\
 & & + J_2\sum_i \vec{s}_{i1}\cdot\vec{s}_{i2}
-\vec{B}\cdot\sum_i(\vec{s}_{i1}+\vec{s}_{i2}), 
\label{1}
\end{eqnarray}
where $<ij>$ runs over the bonds of two $d$-dimensional hypercubes 1 and 2. The 
antiferromagnetically coupled ($J_1>0$) $s=$ {\small $\frac{1}{2}$} Heisenberg 
spins $\vec{s}_{i\eta}$ are coupled locally by $J_2$. Following CM, we first 
integrate out the $J_2$ term\cite{CM,RPA}.
Define the sum and the difference of the spin operators,
\begin{eqnarray}
\vec{S} & = & \vec{s}_1+\vec{s}_2 ,\nonumber \\
\vec{\tilde{S}} & = & \vec{s}_1-\vec{s}_2 ,
\label{2}
\end{eqnarray}
such that
\begin{eqnarray}
{\cal H} & = & \frac{1}{2} J_1 \sum_{<ij>}(\vec{S}_i\cdot\vec{S}_j +\vec{\tilde{S}}_{i}\cdot\vec{\tilde{S}}_{j})+ \nonumber \\
 & & + \frac{1}{4} J_2\sum_i \left( \vec{S}^2_i-\vec{\tilde{S}}^2_i \right)
-\vec{B}\cdot\sum_i \vec{S}_{i}.
\end{eqnarray}
Eq. (\ref{2})
amounts to a transformation to a singlet-triplet basis. Introducing 
hard-core bosons creating the local singlet state, 
$a^{\dagger}_i=\frac{1}{\sqrt{2}}(c^{\dagger}_{i 1\downarrow}c^{\dagger}_{i 2\uparrow}-
c^{\dagger}_{i 1\uparrow}c^{\dagger}_{i 2\downarrow})$, and the local triplet 
$b^{\dagger}_{i 
\;1,0,-1}$ ($b_{1i}^{\dagger}=c^{\dagger}_{i 1\uparrow}c^{\dagger}_{i 
2\uparrow}$ etc.), Eq. 
(\ref{2}) can be alternatively written as,
\begin{eqnarray}
S^z & = & b_1^{\dagger}b_1- b_{-1}^{\dagger}b_{-1}, \nonumber \\
S^+ & = & \sqrt{2}\left(b^{\dagger}_1b_0+b_0^{\dagger}b_{-1} \right), \nonumber \\
\tilde{S}^z & = & -a^{\dagger}b_0-b_0^{\dagger}a, \nonumber \\
\tilde{S}^+ & = & \sqrt{2}\left(b^{\dagger}_1a-a^{\dagger}b_{-1} \right).
\label{3}
\end{eqnarray}
$\vec{S}$ describes $S=1$ spins, while $\vec{\tilde{S}}$ is related to 
fluctuations 
from triplets to singlets. These operators form an $o(4)$ dynamical algebra,
\begin{eqnarray}
{[} S^a,S^b {]} & = & \imath \varepsilon_{abc} S^c ,\label{4a} \\
{[} \tilde{S}^a,\tilde{S}^b {]} & = & \imath \varepsilon_{abc} S^c ,\label{4b} \\
{[} S^a,\tilde{S}^b {]} & = & \imath \varepsilon_{abc} \tilde{S}^c .\label{4c}
\end{eqnarray}
At the $J_2=0$ point (two decoupled layers) the problem has an $O(4)$ global 
invariance, which is broken for any finite $J_2$, leaving only an 
invariance under the $SU(2)$ subgroup Eq. (\ref{4a}).  
The unconventional aspect 
of this problem is that for positive $J_2$ the spontaneous symmetry breaking 
involves the generators $\tilde{S}$. The  $J_2>0$ classical saddle-point of CM is 
easily seen to correspond to the vacuum amplitudes ($z=2d$),
\begin{eqnarray}
\vec{\tilde{\Omega}} & = & \frac{1}{N}\langle \sum_i (-1)^i \vec{\tilde{S}}_i \rangle  = 
\sqrt{1-\frac{J_2^2}{J_1^2z^2}} \Theta\left(J_1 z-J_2\right) \hat{n} 
,\label{5a} \\
n_A & = & \frac{1}{N} \langle \sum_i a^{\dagger}_i a_i \rangle = 
\frac{1}{2}\left(1+\frac{J_2}{J_1z}\right) 
 \Theta\left(J_1 z-J_2\right)  + \nonumber \\
 & &  \hspace{0.17 \hsize} + \Theta\left(J_2-J_1 z \right) ,\label{5b} 
\end{eqnarray}
where $\hat{n}$ is a vector on the unit sphere.
The condensation of $\vec{\tilde{S}}$ (Eq. \ref{5a}) and the existence of a mean singlet 
density (Eq. \ref{5b}) is a direct ramification of the explicit symmetry 
breaking due to the interplanar coupling. $\vec{\tilde{\Omega}}$ is still a vector order parameter, because $\vec{\tilde{S}}$ transforms as a vector 
under $\vec{S}$. It is therefore a N\'eel order state, albeit one with a 
variable local moment size, which implies that its long wavelength behavior 
should be described by the QNLS.

On the classical level it is found that $n_A$ is nonzero for all positive $J_2$, while $\vec{\tilde{\Omega}}$ vanishes continuously at $J_2=J_1 z$, where $n_A$ 
becomes equal to one. This is the transition found by CM. 
Regarding its formal status, it is easily seen that this classical theory 
{\em becomes exact in infinite dimensions} \cite{RPA,Grsea}. 
The energy fluctuations disappear in this limit: $
\Delta E /E \propto 1/\sqrt{N d}$. In addition, we note that
$\vec{\tilde{\Omega}}$ also exists in 2+1 D, at least in the
vicinity of the quantum critical point: the correlation functions in terms of $\vec{s}_1$ and 
$\vec{s}_2$ go to zero at the transition with their ratios fixed according to Eq. (\ref{5a}) \cite{QMC}.

What is wrong with the assertion that this transition and the quantum-critical 
transition in 
2+1 D are the same? The transition in infinite dimensions is a {\em 
classical} transition. In terms of the singlet-triplet 
basis, the quantum fluctuations disappear at the lattice cut-off and thermal 
fluctuations 
dominate at any finite temperature. The numerical study shows quantum 
criticality\cite{QMC}: at zero 
temperature, the quantum fluctuations are scale independent. In the remainder we
will show that this classical theory becomes pathological in the neighborhood
of the classical transition. 

Coherent state path integrals offer a convenient framework to study quantum
order parameter fluctuations
\cite{Fr&Au}. Because of the special status of the order parameter 
Eq. (\ref{5a}), the usual generalized spin coherent states do not suffice. Our 
key result is the 
discovery of a special coherent state for this type of order parameter 
structure. Next to the 
general requirements of normalizability and the existence of the identity, it 
should be 
demanded from coherent states that they reproduce all properties of the 
classical sector. 
Besides reproducing Eq.'s (\ref{5a},\ref{5b}), they should also allow for an 
$\vec{S}$ derived vacuum expectation value,
\begin{equation}
\vec{\Omega}=\frac{1}{N}\langle \sum_i \vec{S}_i \rangle .
\label{7}
\end{equation}
We find that the following coherent state satisfies all these requirements.
\begin{equation}
|\Omega\tilde{\Omega}\rangle = {\rm e}^{\imath \phi S^z} {\rm e}^{\imath \theta 
S^y} {\rm e}^{\imath \theta_2 S^x} 
{\rm e}^{\imath \psi \tilde{S}^y}|\chi\rangle,
\label{8}
\end{equation}
with the reference state,
\begin{equation}
|\chi\rangle=(\cos\chi a^{\dagger}+\sin\chi b_0^{\dagger})|{\rm vac}\rangle .
\label{9}
\end{equation}

Eq. (\ref{8}) looks conventional. It refers to the various rotations
related to the $O$(4) symmetry. The novelty is Eq. (\ref{9}): instead 
of the usual maximum weight state, this non-exact state underlies the
order parameter structure Eq.'s (\ref{5a},\ref{5b}), with $\vec{\tilde{\Omega}}$ chosen
along the $z$ axis, while $\chi$ is fixed by the explicit symmetry breaking
interaction $\sim J_2$. The freedom implied by Eq. (\ref{8}) might at first
instance appear as redundant. However, it turns out that the stiffness
in the temporal direction is caused entirely by the fluctuations from
$\vec{\tilde{\Omega}}$ into the $\vec{\Omega}$ direction, and the four angles appearing
in Eq. (\ref{9}) take care of the independent rotations of $\vec{\tilde{\Omega}}$ and
$\vec{\Omega}$. Explicitely,  $\psi$ parametrizes a rotation from $\vec{\tilde{\Omega}}$ 
to  $\vec{\Omega}\perp\vec{\tilde{\Omega}}$ ($\vec{S}\cdot\vec{\tilde{S}}=0$).
The rotation of $\vec{\tilde{\Omega}}$ in the plane perpendicular to $\vec{\Omega}$ is 
parametrized by $\theta_2$. This is the only free rotation left to $\vec{\tilde{\Omega}}$ in a magnetic field. $\theta$ and $\phi$ fix
the direction of $\vec{\Omega}$. 

We obtain the following expressions for the vacuum amplitudes with respect to 
this coherent state
\begin{eqnarray}
n_A & = & \cos^2\chi\cos^2\psi \label{10} \\
\vec{\Omega} & = & \sin 
2\chi\sin\psi(-\cos\theta\cos\phi,-\cos\theta\sin\phi,\sin\theta) ,
\label{10a} \\
\vec{\tilde{\Omega}} & = & \sin 2\chi \cos\psi\left[\cos\theta_2 \hat{\theta}(\theta,\phi) -\sin\theta_2 \hat{\phi}(\phi) \right],
\label{10b} 
\end{eqnarray}
where $\hat{\theta}$ and $\hat{\phi}$ are the local unit vectors in the $\theta$ and $\phi$ direction, 
$\hat{\theta}  =  (\sin\theta\cos\phi,\sin\theta\sin\phi,\cos\theta)$ and $
\hat{\phi} = (\sin\phi,-\cos\phi,0)$. The identity becomes,  
\begin{eqnarray}
{\large 1}  & = & \int {\rm d} 
\mu\left(\vec{\Omega},\vec{\tilde{\Omega}}\right)|
\Omega\tilde{\Omega}\rangle\langle\Omega\tilde{\Omega}| 
\label{12} \\
 & = &  
 \frac{2}{\pi^4}\int_0^{\pi/2}{\rm d}\chi\int_0^{\pi/2}{\rm 
d}\psi\int_0^{2\pi}{\rm d}\theta_2
\times \nonumber \\
 & & \times
\int_0^{2\pi}{\rm d}\phi\int_{-\pi/2}^{\pi/2}{\rm d}\theta\cos\theta 
|\Omega\tilde{\Omega}\rangle\langle\Omega\tilde{\Omega}|
\nonumber
\end{eqnarray}

By taking expectation values with regard to  $|\Omega\tilde{\Omega}\rangle$ 
(classical
limit), we find the $O(3)$ invariant version of the mean-field theory
of Chubukov and Morr. Minimization of the classical energy with regard to
the coherent state angles yields, 
\begin{eqnarray}
\cos 2\chi_0 & = & \frac{J_2}{J_1 z}+ {\cal O}(B^2),  \label{xmf} \\
\sin\psi_0 & = & \frac{B}{J_1 z} \sqrt{\frac{J_1 z-J_2}{J_1 z+J_2}}
+ {\cal O}(B^2) ,\label{pmf} 
\end{eqnarray} 
with $\theta$ and $\phi$ fixed such that $\vec{\Omega}$ points in the direction 
of 
the magnetic field. We recover the 
classical order-disorder transition at $J_2=J_1z$, where both
$\tilde{\Omega}$ and the induced magnetization $\Omega$ vanish according to 
Eq.'s
(\ref{10a},\ref{10b}). 

The derivation of the path integral is standard\cite{Fr&Au}.  
Using the Trotter formula, the evolution operator in imaginary time
is written as ($N_t$ is the number of time slices, $\delta_t$ the imaginary
time interval, $N_t\delta_t=\beta$),
\begin{equation}
Z=\lim_{\tiny\begin{array}{cc} N_t\rightarrow\infty \\ \delta_t\rightarrow 
0\end{array}} \left({\rm e}^{-\delta_t{\cal H}}\right)^{N_t}.
\label{14}
\end{equation}
Inserting the identity (\ref{12}) at every intermediary time and expanding 
$Z$ to lowest order in $\delta_t$,
\begin{eqnarray}
Z & = & \lim_{\tiny \begin{array}{cc} N_t\rightarrow\infty \\ 
\delta_t\rightarrow 0\end{array}} 
\int {\cal D}\mu \prod_{l=1}^{N_t}
\left[\langle\Omega\tilde{\Omega} (t_l)|\Omega\tilde{\Omega} (t_{l+1})\rangle \right.
\nonumber \\
& & \left.
-\delta_t\langle\Omega\tilde{\Omega} (t_l)|{\cal H}|\Omega\tilde{\Omega} (t_l)\rangle\right],
\end{eqnarray}
where the integration measure ${\cal D}\mu$ is given by $\prod_{l=1}^{N_t} {\rm 
d}\mu\left(\vec{\Omega}_l,\vec{\tilde{\Omega}}_l\right)$, 
while $\{ t_l\}$ is the set of intermediary times in the imaginary time interval [$0,\beta$]. The kinetic term in the action follows from the first term inside
the square brackets,
\begin{eqnarray}
\langle\Omega\tilde{\Omega} (t_l)|\Omega\tilde{\Omega} (t_{l+1})\rangle & = & 
\left(1+{\cal O}(\delta_t^2)\right) {\rm e}^{\imath \delta_t \Phi(t_l)
+{\cal O}(\delta_t^2)},
\label{16}
\end{eqnarray}
with
\begin{eqnarray}
 \Phi & = & \sum_i\sin 2\chi_i\sin\psi_i 
(\sin\theta_l\partial_t\phi_i+\partial_t\theta_{2\,i} ) \nonumber \\
 & = & -\sum_i \sin 2\chi_i \vec{O_i}\cdot \partial_t
\frac{\vec{\tilde{O}}_i}{\tilde{O}_i}\times 
\frac{\vec{\tilde{O}}_i}{\tilde{O}_i} ,
\label{17}
\end{eqnarray}
where $\vec{O}=\vec{\Omega}/\sin 2\chi $ and $\vec{\tilde{O}}=\vec{\tilde{\Omega}}/
\sin 2\chi$, so $O^2+\tilde{O}^2=1$.
 
The potential energy is ($B=0$),
\begin{eqnarray} 
{\cal V} & = & \frac{J_1}{2}\sum_{<i,j>}
\sin 2\chi_i\sin 2\chi_j \left( \vec{O}_i\cdot\vec{O}_j
+  \vec{\tilde{O}}_i\cdot\vec{\tilde{O}}_j \right) + \nonumber \\
 & & + \frac{1}{4}\sum_i\left(1-4 \tilde{O}_i^2 \cos^2\chi_i\right).
\end{eqnarray}

Taking the time continuum limit, the path-integral becomes
$Z=\lim_{\tiny \begin{array}{cc} N_t\rightarrow\infty \\ \delta_t\rightarrow 0\end{array}} 
\int {\cal D}\mu {\rm e}^{-{\cal S}_M }$,
with the real-time action,
\begin{eqnarray}
{\cal S}_M = \int_0^T {\rm d} x_0 \left[ -\Phi(x_0)+{\cal V}(x_0) \right]
\end{eqnarray}

To derive the long wavelength theory,
$\vec{O}$ and $\vec{\tilde{O}}$ are separated
into a slowly varying order-parameter part and a rapidly fluctuating part which 
will be integrated out.  The fluctuations in $\chi$ are massive because of
the explicit symmetry breaking, and can be neglected. We are left with, 
\begin{eqnarray}
\vec{\tilde{O}}_i & = & \eta_i\left(\vec{\tilde{m}}_i+a 
\vec{\tilde{L}}_{\parallel i} \right)
 + a \vec{\tilde{L}}_{\perp i} ,\\
\vec{O}_i & = & \vec{m}_i+ a \vec{L}_i .
\end{eqnarray}
The (staggered) fluctuation $\vec{\tilde{L}}_{\parallel i}$ is parallel to the 
order parameter $\vec{\tilde{m}}_i$ ($\eta_i = \pm 1$ depending
on the sublattice). $\vec{L}$ has a component along $\vec{m}$, but is 
perpendicular
to $\vec{\tilde{m}}$ because of the constraint 
$\vec{O}_i\cdot\vec{\tilde{O}}_i=0$. 
As we already indicated, despite the fact that
the order-parameter part of $\vec{O}$ is zero in the absence of
a magnetic field, the fluctuations in this quantity are actually producing
the stiffness in the time direction and should be carefully integrated out.
We expand to second order in the lattice 
constant $a$, which will be taken to zero at the
end of the calculation. Using the constraint $O_i^2+\tilde{O}_i^2=1$,  the
fluctuation $\vec{\tilde{L}}_{\parallel i}$ is eliminated from the action.
Different from the single-layer system,  
two canting-fields result, $\vec{\tilde{L}}_{\perp i}$ and $\vec{L}_i$,
which have to be integrated out. The former does not influence the
long wavelength behavior, while the
latter, which is related to the response of the system to a uniform magnetic 
field applied perpendicular to the plane of ordering, is responsible for the 
stiffness in the time direction.

After expanding in $a$ and eliminating $\vec{\tilde{L}}_{\parallel}$,
the kinetic term becomes
\begin{eqnarray}
\Phi & = &   -\sum_i\sin 2\chi_i \frac{a}{\tilde{m}_i^2}
\vec{L}_i\cdot\partial_0 \vec{\tilde{m}}_i \times \vec{\tilde{m}}_i
 + \mbox{stagg. terms}. 
\end{eqnarray} 
Using $\vec{\tilde{m}}_i-\vec{\tilde{m}}_{j} \simeq a \partial_{i\rightarrow j} 
\vec{\tilde{m}}_i$,
it can be seen that the staggered terms give contributions which are 
of third order in $a$. The expression for $\Phi$ is identical to that for the
single-layer system, apart from the factor $\sin 2\chi$ and the {\em 
absence of a topological term.} Within the limitations of the semiclassical
expansion, the above derivation is in principle valid
for any dimension, including the 1+1 dimensional two leg spin ladder systems. 
The usual argument for the irrelevance of topological terms in these systems
are based on the proximity of N\'eel order on both chains separately:     
the topological terms in the two rows cancel each other. Here we find 
that this holds regardless the strength of the local fluctuations. We
notice that, according to Haldane's conjecture\cite{Haldane},
the spectrum of the two leg
ladder has to be gapped for any $J_2 \neq 0$. 

The potential term is written in the form
\begin{eqnarray}
 {\cal V} & = & J_1 \sum_{<i,j>}\sin 2\chi_i\sin 2\chi_j \left[ 
\frac{a^2}{4}\left(\partial_{i\rightarrow j} \vec{\tilde{m}}_i\right)^2 +\right. \nonumber \\
 & & \left. +a^2\left( L_i^2+\tilde{L}_{\perp i}^2\right)-2\right]
+J_2\sum_i a^2 L_i^2\cos^2\chi_i \nonumber \\
 & & - \frac{J_2}{4} \sum_i \left(1-4\cos^2\chi_i \right). \label{19}
\end{eqnarray}
In the continuum limit ($a\rightarrow 0$), the summations over sites 
are replaced by integrations over space, 
$\sum_i \rightarrow a^{-d}\int {\rm d}^d x.$ 
The ${\cal O}(1)$ term in Eq. (\ref{19}), corresponding with the mean field 
energy for the bilayer model, acquires a large prefactor $a^{-d}$ and can be 
integrated by steepest descent. This yields the mean-field expression for 
$\chi$,
Eq. (\ref{xmf}).

After integrating over the fluctuations $\vec{L}$ and $\vec{\tilde{L}}_{\perp}$
we recover the effective action responsible for the long wavelength 
fluctuations, which is the  O(3) QNLS

\begin{eqnarray}
{\cal S}_M & = & \frac{1}{2}\int {\rm d}^{d+1}x \left[\chi_{\perp} \left(\partial_0 \vec{\tilde{m}}\right)^2
 - \rho_s \sum_{\alpha=1}^d\left(\partial_{\alpha} \vec{\tilde{m}}\right)^2 \right].
\label{19a}
\end{eqnarray}  
Although the form of Eq. (\ref{19a}) is dictated by symmetry, the parameters
appearing in the effective theory have a quite different meaning in terms
of the microscopic model than is the case in single layer problems. Taking
the saddle point values, 
the perpendicular susceptibility and the spin stiffness become respectively,

\begin{eqnarray}
\chi_{\perp} & = & a^{-d} \frac{J_1 z-J_2}{J_1^2 z^2} ,\\
\rho_s & = & a^{2-d}\frac{J_1}{2}\left(1-\frac{J_2^2}{J_1^2z^2}\right) 
\label{19b}
\end{eqnarray}

{\em Both the susceptibility and the spin stiffness vanish at the classical 
transition at $J_2 = zJ_1$}. 
The spin-wave velocity $v_s=\sqrt{\rho_s/\chi_{\perp}}$ 
remains finite at the transition, 
and no divergencies occur on the Gaussian level\cite{CM}. The stability of the 
classical state against quantum melting is, however, controlled by the 
dimensionless coupling constant $u=a^{1-d}/\sqrt{\rho_s\chi_{\perp}}$ which is 
found to
{\em diverge} at the classical transition as $u \sim 1 / J_2^*$, 
where $J_2^*$ is the reduced interlayer coupling $J_2^* = (J_1 z - J_2 ) / J_2$.
In any finite dimension, the $O(3)$ QNLS quantum critical transition occurs at 
a finite value of the coupling constant and it follows that the long wavelength
fluctuations destroy the $\tilde{\Omega}$ type N\'eel order before the classical critical
point is reached. Accordingly, the quantum critical transition of the bilayer
model is of the $O(3)$ QNLS kind, and the classical transition of Chubukov and
Morr only exists in infinite dimensions.

This theory is even quantitatively reasonable.
One loop renormalization theory for the QNLS in 2+1 
dimensions puts the critical coupling at $u^*=4\pi$ \cite{CHN&CSY}. Using the
saddle point values for the spin stiffness and susceptibility (Eq. \ref{19b}), 
we find the quantum transition to occur at $J_2^c/J_1=3.3$. Given that 
$1/S$-like corrections are neglected\cite{RPA}, the agreement with the value of 
2.5-2.6 obtained
from Quantum Monte Carlo\cite{QMC} and series expansions\cite{Hida1} is
reasonable.

In summary, we have clarified the origin of the quantum critical transition
of the bilayer Heisenberg problem. The key aspect is that the order parameter
structure as discovered by Chubukov and Morr is unusual. Although this order
parameter is macroscopically of the usual $O$(3) vector kind, and therefore
described by the $O$(3) quantum non-linear sigma model, its microscopic status
is unconventional. The operators acquiring a vacuum amplitude ($\tilde{S}$) are 
not the ones
expressing the global $SU$(2) invariance of the problem. This kind of order
parameter structure arises naturally in the present context and we expect it
to be quite common in the general context of quantum magnetism\cite{siti}. 
Our main result is
the discovery of a new type of spin coherent state which allows for the
requantization of such order parameter structures. As applied to the bilayer
problem, the novelty is that in any finite dimension the classical theory
becomes highly pathological: the bare coupling constant of the field theory
diverges at the classical transition, explaining why the
quantum transition obeys O(3) QNLS universality.

{\em Acknowledgments.} Financial support was provided by the Foundation
of Fundamental Research on Matter (FOM), which is sponsored by the
Netherlands Organization for the Advancement of Pure Research (NWO),
and by the Dutch Academy of Sciences (KNAW).


\end{document}